\documentclass[12pt]{article}
\DeclareMathAlphabet{\scr}{U}{rsfs}{m}{n}
\usepackage{latexsym}
\usepackage[mathscr]{eucal}
\usepackage{amsfonts}
\usepackage{amscd}
\usepackage{cite}
\usepackage{amssymb}
\usepackage{colordvi}
\usepackage[centertags]{amsmath}
\usepackage{dsfont}
\usepackage{yfonts}
\usepackage{enumerate}
\usepackage{graphicx}
\usepackage{rotating}

\newcommand{\cleqn}{\setcounter{equation}{0}}
\newcommand{\newc}{\newcommand}
\newc{\be}{\begin{equation}}
\newc{\ee}{\end{equation}}
\newc{\bea}{\begin{eqnarray}}
\newc{\eea}{\end{eqnarray}}
\newc{\ol}{\overline}
\newc{\wt}{\widetilde}
\newc{\bs}{\boldsymbol}
\newc{\m}{\mathcal}
\newcommand{\barr}{\begin{array}}
\newcommand{\earr}{\end{array}}

\begin{document}

\title{\hfill ~\\[-30mm]
       \hfill\mbox{\small SHEP-10-34}\\[30mm]
       \textbf{Spontaneous breaking of ${\bs{SU(3)}}$ to finite family
         symmetries --  a pedestrian's approach}}
\date{}
\author{\\[5mm]Christoph Luhn
\footnote{E-mail: {\tt
      christoph.luhn@soton.ac.uk}}
\\[7mm]
  \emph{\small{
School of Physics and Astronomy, University of Southampton,}}\\
  \emph{\small{Southampton, SO17 1BJ, United Kingdom}}\\
}
\maketitle

\begin{abstract}
\noindent Non-Abelian discrete family symmetries play a pivotal role in
the formulation of models with tri-bimaximal lepton mixing. We discuss how
to obtain symmetries such as $\m A_4$, $\m Z_7 \rtimes \m Z_3$ and
$\Delta(27)$ from an underlying $SU(3)$ gauge symmetry. Higher irreducible
representations are required to achieve the spontaneous breaking of the
continuous group. We present methods of identifying the required vacuum
alignments and discuss in detail the symmetry breaking potentials.
\end{abstract}
\thispagestyle{empty}
\vfill
\newpage
\setcounter{page}{1}

\section{Introduction}
\cleqn

The Standard Model of particle physics provides a successful and accurate
description of Nature as has been proved in countless experiments over the
last few decades. Yet, the observation of neutrino oscillations demands its
extension to include massive neutrinos. Due to our ignorance of the absolute
neutrino mass scale, the structure of the neutrino mass spectrum is still in
the dark with hierarchical and quasi-degenerate scenarios being equally well 
conceivable. A better clue towards understanding the underlying physics of
flavor is given by the observed mixing pattern in the lepton sector. While the
quarks mix with three small angles, the lepton mixing features one small and
two large angles. Even more intriguing is the fact that the best fit values
\cite{Schwetz:2008er,GonzalezGarcia:2010er} for the lepton mixing angles are
remarkably close to the so-called tri-bimaximal pattern
\cite{Harrison:2002er,Harrison:2002kp}, 
\be
\begin{pmatrix} 
-\frac{2}{\sqrt{6}} & ~\frac{1}{\sqrt{3}} & 0 \\
\phantom{-}\frac{1}{\sqrt{6}} & ~\frac{1}{\sqrt{3}} &  \phantom{-}\frac{1}{\sqrt{2}}\\
\phantom{-}\frac{1}{\sqrt{6}} & ~\frac{1}{\sqrt{3}} &  -\frac{1}{\sqrt{2}}
\end{pmatrix},
\ee
corresponding to $\theta_{12} = 35.26^\circ$, $\theta_{23} = 45^\circ$,
$\theta_{13} = 0^\circ$. This peculiar mixing pattern suggests a non-Abelian
discrete family symmetry $\m G$ lurking behind the flavor structure of the
chiral fermions. The virtue of imposing such a non-Abelian symmetry is that
the irreducible representations (irreps) of $\m G$ allow one to collect the
families of chiral fermions into multiplets. With three known families it is
natural to investigate finite groups with triplet and/or doublet
representations. These are found among the finite subgroups of $SU(3)$,
$SU(2)$ and $SO(3)$, with popular candidates being $\m A_4$, $\m S_4$ and
$\Delta(27)$. Adopting their preferred finite group, many authors have
constructed even more models of flavor, all aiming to explain the remarkable
tri-bimaximal mixing pattern. We refer the reader to the review by Altarelli
and Feruglio \cite{Altarelli:2010gt} which includes an extensive list of
references of such models.

In this paper we wish to address questions relating to a possible gauge origin
of the non-Abelian discrete family symmetry. A symmetry $\m G$ is called a
{\it discrete gauge symmetry} if it originates from a spontaneously
broken gauge symmetry $G$. The assumption of a gauge origin has the advantage
that the remnant discrete symmetry $\m G$ is protected against violations by
quantum gravity effects \cite{Krauss:1988zc}. 

This idea has been applied to Abelian symmetries
\cite{Ibanez:1991hv,Ibanez:1991pr,Dreiner:2005rd,Luhn:2007gq} and is well
established and understood. Assuming a gauged $U(1)$ symmetry with integer
charge normalization, one obtains a residual $Z_N$ symmetry when a field
$\phi$ with $U(1)$ charge $N$ develops a vacuum expectation value (VEV) via a
potential of the form
\vspace{-0.5mm}
\be
V ~=~ -m^2 \phi^\dagger \phi + \lambda (\phi^\dagger \phi)^2 \ .
\ee
\vspace{-6.5mm}

\noindent The resulting would-be Goldstone boson of the spontaneously broken
$U(1)$ symmetry is then eaten by the $U(1)$ gauge boson's longitudinal
polarization. 

The situation is much more involved in the non-Abelian case since higher
representations of the continuous gauge group $G$ are required to achieve the
desired breaking. The breaking patters of $G=SO(3)$ using low-dimensional
representations have been investigated in
\cite{Ovrut:1977cn,Etesi:1997jv,Koca:1997td,Koca:2003jy,Berger:2009tt}. In the
context of flavor models, the most interesting result of these studies is that
the tetrahedral group $\m A_4$ can originate from an $SO(3)$ symmetric potential
involving only the ${\bf 7}$ representation. The free parameters of the
potential can be chosen without fine-tuning so that the potential is minimized
by a VEV which breaks $SO(3)$ but not $\m A_4$.

It is the purpose of this paper to similarly examine the case of $G=SU(3)$. A
first attempt in this direction has been undertaken in
\cite{Adulpravitchai:2009kd} where the $SU(3)$ representations ${\bf 3}$,
${\bf 6}$ and ${\bf 8}$ have been considered to achieve the breaking of the
continuous symmetry. It is shown there that these small representations are
insufficient to generate a remnant discrete symmetry with triplet
representations like e.g. $\m A_4$. Furthermore, the study stops short of
discussing the potential and the relevant order parameters that determine the
breaking of $SU(3)$ to the discrete symmetry $\m G$. In the present work we go
beyond \cite{Adulpravitchai:2009kd} by (a) discussing also higher
representations of $SU(3)$ and (b) scrutinizing the relevant symmetry breaking
potential.

The paper is structured as follows. In section~\ref{sec-dec} we present a
simple way to identify the embedding of a given finite group $\m G$ in
$SU(3)$. Having worked out the decomposition of $SU(3)$ representations under
$\m G$, we discuss the procedure of finding the $\m G$ singlet directions of
the appropriate $SU(3)$ irreps in section~\ref{sec-find}. Along the way we
also comment on the choice of basis of the finite subgroup. In
section~\ref{sec-subgroup} we work out the maximal subgroup that is left
invariant by a VEV in such a singlet direction. Section~\ref{sec-pot} is
devoted to the study of several symmetry breaking potential which can give
rise to $\m A_4$, $\m Z_7\rtimes \m Z_3$ and $\Delta(27)$, respectively.
Finally, we conclude in section~\ref{sec-conclusion}.




\section{\label{sec-dec}Decomposition of $\bs{SU(3)}$ irreps}
\cleqn

In order to break $SU(3)$ spontaneously down to a finite subgroup $\m G$ it
is necessary to find those $SU(3)$ irreps which contain a singlet of $\m G$ in
their decomposition. A simple method for obtaining the full decompositions
is based on the observation that all $SU(3)$ irreps ${\bs{\rho}}$ can be
successively generated from the fundamental ${\bf 3}$. The complex conjugate
representations $\ol{\bs{\rho}}$ are directly derived from ${\bs{\rho}}$.
Table~\ref{tab-su3} lists the relevant tensor products that can be used
to find the irreps up to dimension 27. The last number in each line shows the
new irrep that is generated from multiplying already known ones. 

%
%
\begin{table}
$$
\begin{array}{rcl}
\multicolumn{3}{c}{\phantom{\Big|}\text{some~}SU(3)~\text{tensor~products}}
\phantom{\Big|}\\\hline\hline
\phantom{\Big|}{\bf 3\otimes 3} &\!=\!& {\bf \ol 3 + 6}\phantom{\Big|}\\
\phantom{\Big|}{\bf 3\otimes \ol 3} &\!=\!& {\bf 1 + 8}\phantom{\Big|}\\
\phantom{\Big|}{\bf 6\otimes 3} &\!=\!& {\bf 8 + 10}\phantom{\Big|}\\
\phantom{\Big|}{\bf 6 \otimes \ol 3} &\!=\!& {\bf 3 + 15}\phantom{\Big|}\\
\phantom{\Big|}{\bf 10 \otimes 3} &\!=\!& {\bf 15 + 15'}\phantom{\Big|}\\
\phantom{\Big|}{\bf \ol{10}\otimes 3} &\!=\!& {\bf \ol 6 + 24}\phantom{\Big|}\\
\phantom{\Big|}{\bf \ol{10} \otimes \ol{6}} &\!=\!& {\bf 15 + 24 + 21}\phantom{\Big|}\\
\phantom{\Big|}{\bf 6 \otimes \ol 6} &\!=\!& {\bf 1 + 8 +
  27}\phantom{\Big|}
\end{array}
$$
\caption{\label{tab-su3}A list of $SU(3)$ tensor products which can be used to
successively obtain the $SU(3)$ irreps up to dimension 27.}
\end{table}
%
%

Identifying the triplet of $SU(3)$ with a faithful representation of $\m G$,
one can successively work out the decomposition of all ${\bs{\rho}}$ by
comparing the $SU(3)$ tensor products with the Kronecker products of $\m G$.
This method is best illustrated for explicit examples. Let us consider the
case of the tetrahedral group $\m A_4=\Delta(12)$ as well as $\Delta(27)$.

\begin{itemize}
\item[$(i)$] $\m A_4=\Delta(12)$ has four irreps
${\bf 1}, {\bf 1'}, \ol{\bf 1'}$ and the real ${\bf 3}$
which satisfy the following multiplication rules.
\vspace{-4.2mm}
\begin{center}
\begin{tabular}{c}
 \\[-2mm]
{${\m A_4}$  Kronecker products}\hfill\\
   \\[-3mm]
\hline    
 \\[-3mm]
${\bf 1'}\otimes \,{\bf 1'}=~\ol{\bf 1'}\hfill$ \\
${\bf 1'}\otimes \,\ol{\bf 1'}=~{\bf 1}\hfill$ \\
${\bf 3}\:\otimes \,{\bf 1'}=~{\bf 3}\hfill$ \\
${\bf 3}\:\otimes\,{\bf 3}\:=~{\bf 1}+{\bf 1'}+\ol{\bf 1'}+2\cdot{\bf 3}\hfill$ 
\end{tabular}
\end{center}
As the $\m A_4$ triplet is real, we can identify it with both the ${\bf 3}$ as
well as the $\ol{\bf 3}$ of $SU(3)$. Comparing the products of ${\bf 3 \otimes
  3}$ we directly find the decomposition of the sextet, ${\bf 6}
\rightarrow {\bf{1 + 1' + \ol {1'} + 3}}$. The decomposition of the octet is
obtained similarly from ${\bf 3 \otimes \ol  3}$, leading to ${\bf 8}
\rightarrow {\bf{ 1' + \ol {1'}}} +2 \!\cdot\! {\bf 3 }$. For the ${\bf{ 10}}$
we consider the $SU(3)$ tensor product ${\bf 6 \otimes 3} = {\bf
  8+10}$. Plugging in the just determined $\m A_4$ decompositions we find
\vspace{-2mm}
\bea
{\bf 10} &\rightarrow & \underbrace{({\bf{1 + 1' + \ol {1'} + 3}})}_{{\bf 6}}
\otimes \, {\bf{3}} \,-\,
\underbrace{({\bf{ 1' + \ol {1'}}} + 2\!\cdot\! {\bf 3})}_{{\bf 8}}  ~=~ {\bf{ 1 }} + 3\!\cdot\! {\bf
  3}\ ,\notag 
\eea
\vspace{-3.8mm}

\noindent where, in the last step, we have used the $\m A_4$ Kronecker
products. Continuation of these simple calculations yields the 
decomposition of any $SU(3)$ irrep. We list the results up to the ${\bf
  27}$, cf. also \cite{Luhn:2008sa}. 
\vspace{-3.7mm}
\begin{center}
\begin{tabular}{c} 
\\[-3mm]
~~${ SU(3)~ \supset~ \m A_4}$  \hfill\\
   \\[-3mm]
\hline    
 \\[-3mm]
 ${\bf 3}~=~{\bf 3}\hfill$ \\ 
 ${\bf 6}~=~{\bf 1}+{\bf 1'}+\ol{\bf 1'} + {\bf 3}\hfill$ \\
${\bf 8}~=~ {\bf 1'}+\ol{\bf 1'} + 2\cdot {\bf 3} \hfill$ \\ 
${\bf 10}~=~ {\bf 1} + 3\cdot {\bf 3} \hfill$ \\
${\bf 15}~=~  {\bf 1}+{\bf 1'}+\ol{\bf 1'} + 4\cdot{\bf 3} \hfill$ \\
${\bf 15'}\;\!=~ 2\cdot({\bf 1}+{\bf 1'}+\ol{\bf 1'}) + 3\cdot{\bf 3} \hfill$ \\
${\bf 21}~=~ {\bf 1}+{\bf 1'}+\ol{\bf 1'} + 6\cdot{\bf 3}  \hfill$ \\
${\bf 24}~=~ 2\cdot({\bf 1}+{\bf 1'}+\ol{\bf 1'}) + 6\cdot{\bf 3} \hfill$ \\
${\bf 27}~=~ 3\cdot({\bf 1}+{\bf 1'}+\ol{\bf 1'}) + 6\cdot{\bf 3}   \hfill$
\\[-0.1mm] 
\end{tabular}
\end{center}
This shows that the irreps ${\bf 6}$, ${\bf 10}$, ${\bf 15}$,
${\bf 15'}$, ${\bf 21}$, ${\bf 24}$ and ${\bf 27}$ contain at least one
singlet of $\m A_4$ and can thus, in principle, be used to break $SU(3)$
spontaneously down to $\m A_4$ or a group that contains $\m A_4$ as a subgroup.
\item[$(ii)$] $\Delta(27)$ has nine one-dimensional irreps 
$$
\!\begin{array}{lllll}
{\bf 1}={\bf 1_{0,0}}\,,
& \, {\bf 1_1}={\bf 1_{0,1}}\,,&\,{\bf 1_3}={\bf 1_{1,0}}\,,&\,{\bf
  1_5}={\bf 1_{1,1}}\,,&\,{\bf 1_7}={\bf 1_{1,2}}\,, \\[4mm]
& \, {\bf 1_2}={\bf \ol 1_1}={\bf 1_{0,2}}\,,&\,{\bf 1_4}={\bf \ol 1_3}={\bf
  1_{2,0}}\,,&\,{\bf 1_6}={\bf \ol 1_5}={\bf 1_{2,2}}\,,&\,{\bf 1_8}={\bf \ol
  1_7}={\bf 1_{2,1}}\,, 
\end{array}
$$
as well as a triplet ${\bf 3}$ and its complex conjugate $\ol{\bf 3}$. The
Kronecker products read as follows.
\begin{center}
\begin{tabular}{c} 
\\[-3mm]
{${\Delta(27)}$  Kronecker products}\hfill\\
   \\[-3mm]
\hline   
 \\[-2mm]
${\bf 1_{r,s}}\;\!\otimes \,{\bf 1_{r',s'}}\,=~{\bf 1_{r+r',s+s'}}\hfill$ \\
${\bf 3}~\otimes \,{\bf 1_j}\,=~{\bf 3}\hfill$ \\
$\ol{\bf 3}~\otimes \,{\bf 1_j}\,=~\ol{\bf 3}\hfill$ \\
${\bf 3}~\otimes \;{\bf 3}~=~3\cdot \ol{\bf  3}\hfill$ \\
${\bf 3}~\otimes \;\ol{\bf 3}~=~{\bf 1} + \sum_{j=1}^8 {\bf  1_j}\hfill$  \\[2mm]
\end{tabular}
\end{center}
Here $r,s=0,1,2$ and the sums $r+r'$ and $s+s'$ are taken modulo~3.
Without loss of generality we can identify the ${\bf 3}$ of $SU(3)$ with the
${\bf 3}$ of $\Delta(27)$. Then also their complex conjugates automatically
correspond to each another. Comparing the product ${\bf 3\otimes 3}$ gives the
decomposition of the sextet, ${\bf 6} \rightarrow 2\!\cdot\!\ol{\bf{3}}$. From 
${\bf 3}\otimes \ol{\bf 3}$ we derive the decomposition of the octet, ${\bf 8}
\rightarrow \sum_{j=1}^8 {\bf  1_j}$. The ${\bf 10}$ is again obtained from
the $SU(3)$ tensor product ${\bf 6 \otimes 3} = {\bf 8+10}$.
\bea
{\bf 10} &\rightarrow & \underbrace{(2\!\cdot\!\ol{\bf{3}})}_{{\bf 6}} \otimes
\,{\bf{3}} \,-\,
\underbrace{\sum_{j=1}^8 {\bf  1_j}}_{{\bf 8}}  ~=~ 2\!\cdot\!{\bf{ 1 }} + 
\sum_{j=1}^8 {\bf  1_j} \ .\notag 
\eea
Analogously we get the decomposition for any other $SU(3)$ irrep
showing that, for irreps up to dimension 27, only the ${\bf
  10}$ and the ${\bf   27}$ contain singlets of $\Delta(27)$,
cf. \cite{Luhn:2008sa}.   
\begin{center}
\begin{tabular}{c}
\\[-3mm]
~~${ SU(3)~ \supset~ \Delta(27)}$  \hfill\\
   \\[-3mm]
\hline    
 \\[-2mm]
 ${\bf 3}~=~{\bf 3}\hfill$ \\ 
 ${\bf 6}~=~2\cdot \ol{\bf 3}\hfill$ \\
${\bf 8}~=~\sum_{j=1}^8 {\bf 1_j}\hfill$ \\ 
${\bf 10}~=~2\cdot {\bf 1} + ~\sum_{j=1}^8 {\bf 1_j}\hfill$ \\
${\bf 15}~=~5\cdot {\bf 3} \hfill$ \\
${\bf 15'}\;\!=~5\cdot {\bf 3}   \hfill$ \\
${\bf 21}~=~ 7\cdot {\bf 3}  \hfill$ \\
${\bf 24}~=~8\cdot {\bf 3}   \hfill$ \\
${\bf 27}~=~ 3\cdot ({\bf 1} + \sum_{j=1}^8 {\bf 1_j})\hfill$ \\[2mm] 
\end{tabular}
\end{center}
\end{itemize}

The same procedure can be repeated for any other finite subgroup $\m G$ of
$SU(3)$
\cite{MBD,FFK,BLW,Luhn:2007uq,Luhn:2007yr,Escobar:2008vc,Ludl:2009ft,Ishimori:2010au,Grimus:2010ak,ramond-book}. This
way it is possible to identify those irreps which can  potentially break
$SU(3)$ down to~$\m G$. Table~\ref{tab-singlets} summarizes these results by
listing the number of singlets of $\m G$ within each $SU(3)$ irrep for various
finite subgroups. 
\begin{table}
$$
\begin{array}{cccccccccc}
\phantom{\Big|}\text{finite~subgroup~}\m G\phantom{\Big|} & ~{\bf 3}~ & ~{\bf 6}~ &
~{\bf 8}~ & \,{\bf 10}\,& \,{\bf 15}\,& \,{\bf
  15'}\,&\, {\bf 21}\,& \,{\bf 24}\,& \,{\bf 27}\, \\\hline\hline
\phantom{\Big|}\m A_4 = \Delta(12)\phantom{\Big|} &-&1&-&1&1&2&1&2&3 \\
\phantom{\Big|}\Delta(27)\phantom{\Big|} &-&-&-&2&-&-&-&-&3 \\
\phantom{\Big|}\m S_4 = \Delta(24)\phantom{\Big|} &-&1&-&-&-&2&-&1&2 \\
\phantom{\Big|}\Delta(54)\phantom{\Big|} &-&-&-&-&-&-&-&-&3 \\
\phantom{\Big|} \m Z_7 \rtimes \m Z_3  = \m T_7 \phantom{\Big|} &-&-&-&1&1&1&1&1&1 \\
\phantom{\Big|}\m{PSL}_2(7) = \Sigma(168) \phantom{\Big|} &-&-&-&-&-&1&-&-&- \\
\end{array}
$$
\caption{\label{tab-singlets}The number of singlets of $\m G$ within each
  $SU(3)$ irrep for various finite subgroups.}
\end{table}


\section{\label{sec-find}Finding the singlet direction}
\cleqn

In the previous section we have determined the $SU(3)$ irreps that
contain singlets of the finite subgroup $\m G$. The next step is to find the
directions of these representation which correspond to the singlets. It is
worth emphasizing that such singlet VEVs may or may not break $SU(3)$ directly
to the desired finite group $\m G$. In the latter case, a bigger subgroup of
$SU(3)$ is left intact and the breaking to $\m G$ can be achieved {\it
sequentially} by adding a second irrep with an appropriate singlet
VEV.\footnote{An example of such a sequential breaking is discussed in
section~\ref{sec-subgroup}. There we will show that $\m A_4$ cannot be
obtained directly from the ${\bf 6}$ or ${\bf 10}$ alone but only their
combination.} Focusing on the smallest irreps we confine ourselves to the
${\bf 6}$, ${\bf 10}$ and ${\bf 15}$ in the following. We construct them using
the fundamental triplet.   

The three orthonormal states of an $SU(3)$ triplet are denoted by
$|\,i\,\rangle$, with $i=1,2,3$. Then we can express a general triplet as a
linear combination
\be
\sum_{i=1}^3 \varphi_i |\,i\,\rangle \ ,
\ee
with $\varphi_i$ being the components of the state.


The ${\bf 6}$ of $SU(3)$ corresponds to the symmetric product of two
triplets. Using the compact notation $|\,ij\,\rangle \equiv |\,i\,\rangle
\otimes |\,j\,\rangle$ we can define six orthonormal states
$|\,\alpha\,\}$, where $\alpha=1,...,6$, as follows

\be
|\,1 \,\} \,=\: |\,11\,\rangle\ , \qquad
|\,2 \,\} \,=\: |\,22\,\rangle\ , \qquad
|\,3 \,\} \,=\: |\,33\,\rangle\ ,\notag
\ee
\vspace{-2mm}
\be
|\,4 \,\} \,=\: \frac{1}{\sqrt{2}}\left( |\,12\,\rangle + |\,21\,\rangle
\right) , \notag
\ee
\vspace{-1mm}
\be
|\,5 \,\} \,=\: \frac{1}{\sqrt{2}}\left( |\,23\,\rangle + |\,32\,\rangle
\right) , \notag
\ee
\vspace{-1mm}
\be
|\,6 \,\} \,=\: \frac{1}{\sqrt{2}}\left( |\,31\,\rangle + |\,13\,\rangle
\right) . \label{six}
\ee
A general sextet state is then given by
\vspace{-1mm} 
\be
\sum_{\alpha=1}^6\chi_\alpha |\,\alpha\,\} ~=\: \sum_{i,j=1}^3 T_{ij} \,
|\,ij\,\rangle \ , \label{sixtensor}
\ee
\vspace{-2mm}

\noindent
where $\chi_\alpha$ denotes the six independent components of the sextet state
and $T_{ij}$ is the corresponding symmetric tensor. $T_{ij}$ and $\chi_\alpha$
are related via Eqs.~(\ref{six},\ref{sixtensor}). For example, $T_{11} =
\chi_1$ and $T_{12} =T_{21} =\frac{1}{\sqrt{2}} \chi_2$.



The ${\bf 10}$ of $SU(3)$ corresponds to the symmetric product of three
triplets. We can define its orthonormal basis $|\,a \succ$, with $a=1,...,10$,
by 
\be
|\,1 \succ \,\,=\: |\,111\,\rangle\ , \qquad
|\,2 \succ \,\,=\: |\,222\,\rangle\ , \qquad
|\,3 \succ \,\,=\: |\,333\,\rangle\ ,\notag
\ee
\vspace{-2mm}
\be
|\,4 \succ \,\,=\, \frac{1}{\sqrt{3}}\left(|\,112\,\rangle +  |\,121\,\rangle
  + |\,211\,\rangle \right) , \quad~
|\,5 \succ \,\,=\, \frac{1}{\sqrt{3}}\left(|\,113\,\rangle +  |\,131\,\rangle
  + |\,311\,\rangle \right) ,\notag
\ee
\vspace{-1mm}
\be
|\,6 \succ \,\,=\, \frac{1}{\sqrt{3}}\left(|\,221\,\rangle +  |\,212\,\rangle
  + |\,122\,\rangle \right) , \quad~
|\,7 \succ \,\,=\, \frac{1}{\sqrt{3}}\left(|\,223\,\rangle +  |\,232\,\rangle
  + |\,322\,\rangle \right) ,\notag
\ee
\vspace{-1mm}
\be
|\,8 \succ \,\,=\, \frac{1}{\sqrt{3}}\left(|\,331\,\rangle +  |\,313\,\rangle
  + |\,133\,\rangle \right) , \quad~
|\,9 \succ \,\,=\, \frac{1}{\sqrt{3}}\left(|\,332\,\rangle +  |\,323\,\rangle
  + |\,233\,\rangle \right) ,\notag
\ee
\vspace{-1mm}
\be
|\,10 \succ \,\,=\, \frac{1}{\sqrt{6}}\left(|\,123\,\rangle +  |\,231\,\rangle
  + |\,312\,\rangle +  |\,321\,\rangle +  |\,213\,\rangle
  + |\,132\,\rangle    \right) .\label{ten}
\ee
Again, the most general state reads
\be
\sum_{a=1}^{10}\psi_a |\,a\succ ~= \sum_{i,j,k=1}^3 T_{ijk} \,
|\,ijk\,\rangle \ , \label{tentensor}
\ee
with Eqs.~(\ref{ten},\ref{tentensor}) relating $\psi_a$ and $T_{ijk}$,
e.g. $T_{112}=T_{121}=T_{211}=\frac{1}{\sqrt{3}}\psi_4$.



Turning to the ${\bf 15}$ of $SU(3)$ we define its orthonormal basis $|\,A
\,)$, with $A=1,...,15$, as
\be
|\,1 \,) \,=\, \frac{1}{\sqrt{3}}\left(|\,11\bar 1\,\rangle -  |\,12\bar
  2\,\rangle - |\,21\bar 2\,\rangle \right) , \notag
\ee
\vspace{-1mm}
\be
|\,2 \,) \,=\, \frac{1}{2\sqrt{6}}\left(
2 \cdot |\,11\bar 1\,\rangle  +  |\,12\bar 2\,\rangle + |\,21\bar 2\,\rangle 
- 3 \cdot |\,13\bar 3\,\rangle  - 3 \cdot |\,31\bar 3\,\rangle  \right) , \notag
\ee
\vspace{-1mm}
\be
|\,3 \,) \,=\, \frac{1}{\sqrt{3}}\left(|\,22\bar 2\,\rangle -  |\,23\bar
  3\,\rangle - |\,32\bar 3\,\rangle \right) , \notag
\ee
\vspace{-1mm}
\be
|\,4 \,) \,=\, \frac{1}{2\sqrt{6}}\left(
2 \cdot |\,22\bar 2\,\rangle  +  |\,23\bar 3\,\rangle + |\,32\bar 3\,\rangle 
- 3 \cdot |\,21\bar 1\,\rangle  - 3 \cdot |\,12\bar 1\,\rangle  \right) , \notag
\ee
\vspace{-1mm}
\be
|\,5 \,) \,=\, \frac{1}{\sqrt{3}}\left(|\,33\bar 3\,\rangle -  |\,31\bar
  1\,\rangle - |\,13\bar 1\,\rangle \right) , \notag
\ee
\vspace{-1mm}
\be
|\,6 \,) \,=\, \frac{1}{2\sqrt{6}}\left(
2 \cdot |\,33\bar 3\,\rangle  +  |\,31\bar 1\,\rangle + |\,13\bar 1\,\rangle 
- 3 \cdot |\,32\bar 2\,\rangle  - 3 \cdot |\,23\bar 2\,\rangle  \right) , \notag
\ee
\vspace{-1mm}
\be
|\,7 \,) \,=\, |\,11\bar 2\,\rangle \ , \qquad
|\,8 \,) \,=\, |\,11\bar 3\,\rangle \ , \qquad
|\,9 \,) \,=\, |\,22\bar 3\,\rangle \ , \notag
\ee
\vspace{-2mm}
\be
|\,10 \,) \,=\, |\,22\bar 1\,\rangle \ , \;\: \quad
|\,11 \,) \,=\, |\,33\bar 1\,\rangle \ , \;\: \quad
|\,12 \,) \,=\, |\,33\bar 2\,\rangle \ , ~\notag
\ee
\vspace{-2mm}
\be
|\,13 \,) \,=\, \frac{1}{\sqrt{2}}\left( |\,12\bar 3\,\rangle + |\,21\bar
  3\,\rangle \right) , \notag
\ee
\vspace{-1mm}
\be
|\,14 \,) \,=\, \frac{1}{\sqrt{2}}\left( |\,23\bar 1\,\rangle + |\,32\bar
  1\,\rangle \right) , \notag
\ee
\vspace{-1mm}
\be
|\,15 \,) \,=\, \frac{1}{\sqrt{2}}\left( |\,31\bar 2\,\rangle + |\,13\bar
  2\,\rangle \right) .\label{fifteen}
\ee
The most general state is now given by
\be
\sum_{A=1}^{15} \Sigma_A  |\,A \,) ~= \sum_{i,j,k=1}^3 T_{ij}^k \, |\,ij\bar
k\,\rangle \ . \label{fifteentensor}
\ee
The fifteen independent components $\Sigma_A$ of the ${\bf 15}$ are related to
the tensor $T_{ij}^k$ via Eqs.~(\ref{fifteen},\ref{fifteentensor}),
e.g. $T_{12}^2=T_{21}^2 = - \frac{1}{\sqrt{3}} \Sigma_1 + \frac{1}{2\sqrt{6}}
\Sigma_2$. Note that $T_{ij}^k$ is symmetric in $i,j$ as well as traceless,
i.e. $\sum_{k=1}^3 T_{ik}^k = 0$.


Having defined the $SU(3)$ irreps ${\bs{\rho}}$ in terms of triplets and
anti-triplets, we now have to fix the basis of the triplet generators of the
finite subgroup $\m G$ in order to see which direction of ${\bs{\rho}}$ is
left invariant under $\m G$. A particularly simple basis for the triplets of
$\Delta(3n^2)$, $\Delta(6n^2)$ as well as $\m Z_7 \rtimes \m Z_3$ is based on
the matrices \cite{King:2009ap}
\begin{equation}
D~=~\begin{pmatrix}
e^{i\vartheta_1} &0&0 \\
0& e^{i\vartheta_2}&0 \\
0&0& e^{-i(\vartheta_1+\vartheta_2)}
\end{pmatrix}  , \quad
A~=~\begin{pmatrix} 0&1&0\\0&0&1\\1&0&0\end{pmatrix}  , \quad
B~=~- \begin{pmatrix} 0&0&1 \\0&1&0 \\ 1&0&0\end{pmatrix}  .\label{gen-under}
\end{equation}
The generators of $\Delta(3n^2)$ are given by $A$ and $D$ with $\vartheta_1=0$
and $\vartheta_2={2 \pi l}/{n}$, where $l\in\mathbb N$. Adding the generator $B$
yields the group $\Delta(6n^2)$. The triplet representation of 
$\m Z_7 \rtimes \m Z_3$ can be defined via $A$ and $D$ with
$\vartheta_1=\vartheta_2/2=2\pi/7$. 

In the following we consider the $SU(3)$ irreps ${\bf 6}$, ${\bf 10}$ and
${\bf 15}$ and determine the singlet directions for the respective groups as
shown in Table~\ref{tab-singlets}.

\begin{itemize}
\item Starting with the ${\bf 6}$ as given in Eq.~(\ref{six}) we see that a
  state with $\chi_1=\chi_2=\chi_3$ and $\chi_4=\chi_5=\chi_6=0$ remains
  invariant under $A$, $B$ and $D_{(\vartheta_1=0,\vartheta_2=\pi)}$. Therefore the
  singlet of $\m A_4$ as well as $\m S_4$ within the ${\bf 6}$ of $SU(3)$
  points into the direction
\be
\m A_4 \,,\, \m S_4 ~\text{singlet~within~the~}{\bf 6} :~ \propto \,
(1,1,1,0,0,0)^T \ . \label{vev6}
\ee
\item For the ${\bf 10}$, see Eq.~(\ref{ten}), we can easily identify a
  singlet direction which is common to all groups generated by $A$ and $D$
  with arbitrary angles $\vartheta_i$. It is given by $\psi_a=0$ for $a=1,...,9$,
\be
\m A_4 \,,\, \Delta(27) \,,\, \m Z_7 \rtimes \m Z_3 ~\text{singlet~within~the~}{\bf 10}
:~ \, \propto (0,0,0,0,0,0,0,0,0,1)^T \ . \label{vev10a}
\ee
Additionally, there exists a second $\Delta(27)$ singlet defined by
$\psi_1=\psi_2=\psi_3$ and $\psi_a=0$ for $a=4,...,10$,
\be
\Delta(27) ~\text{singlet~within~the~}{\bf 10} :~ \, \propto (1,1,1,0,0,0,0,0,0,0)^T \ . \label{vev10b}
\ee
\item Finally, the ${\bf 15}$, see Eq.~(\ref{fifteen}), contains a singlet of
  $\m A_4$, given by $\Sigma_{13}=\Sigma_{14}=\Sigma_{15}$ and
  $\Sigma_{A}=0$ for $A=1,...,12$, 
\be
\m A_4 ~\text{singlet~within~the~}{\bf 15}
:~ \, \propto (0,0,0,0,0,0,0,0,0,0,0,0,1,1,1)^T \ .\label{vev15a}
\ee
The $\m Z_7 \rtimes \m Z_3$ singlet is obtained by setting all components of
the fifteen to zero except for $\Sigma_{7}=\Sigma_{9}=\Sigma_{11}$
\be
\m Z_7 \rtimes \m Z_3 ~\text{singlet~within~the~}{\bf 15}
:~ \, \propto (0,0,0,0,0,0,1,0,1,0,1,0,0,0,0)^T \ . \label{vev15b}
\ee
\end{itemize}


\section{\label{sec-subgroup}Unbroken subgroups}
\cleqn

Having obtained the singlet directions of a particular $SU(3)$ irrep with
respect to the finite subgroup $\m G$, the question arises whether  a
VEV in this particular direction breaks $SU(3)$ down to $\m G$ or some bigger
subgroup. For instance, from Eq.~(\ref{vev6}) we already see that the given
VEV is invariant not only under $\m A_4$ but also $\m S_4$. We will argue in a
moment that such a VEV actually leaves an even bigger group unbroken. To see
this let us parameterize a general $SU(3)$ transformation $U$ in the standard
way 
\be
U~=~ P_1 \cdot
\begin{pmatrix} 
c_{12}c_{13} & s_{12}c_{13} & s_{13}e^{-i\delta}\\
-s_{12}c_{23}-c_{12}s_{23}s_{13}e^{i\delta} &
c_{12}c_{23}-s_{12}s_{23}s_{13}e^{i\delta} & s_{23}c_{13}\\
s_{12}s_{23}-c_{12}c_{23}s_{13}e^{i\delta} &
-c_{12}s_{23}-s_{12}c_{23}s_{13}e^{i\delta} & c_{23}c_{13}
\end{pmatrix}
\cdot P_2 \ ,\label{su3trafo}
\ee
where $c_{ij}=\cos \theta_{ij}$ and $s_{ij}=\sin \theta_{ij}$. In addition to
the three angles $\theta_{ij}$ there are five phases: $\delta$ as well as
$\alpha_i$ and $\beta_i$ as given in the phase matrices 
\be
P_1 ~=~\begin{pmatrix} e^{i\alpha_1} &0&0 \\ 0&e^{i\alpha_2} &0 \\0&
  0&e^{-i(\alpha_1+\alpha_2)}\end{pmatrix}  ,\qquad 
P_2 ~=~\begin{pmatrix}e^{i\beta_1} &0&0 \\ 0&e^{i\beta_2} &0 \\0&
  0&e^{-i(\beta_1+\beta_2)} \end{pmatrix} .
\ee
A general $SU(3)$ transformation of a triplet state $|\,i\,\rangle$  now
takes the form 
\be
|\,i\,\rangle ~\rightarrow~ \sum_{j=1}^3 U_{ij}\, |\,j\,\rangle \ .
\ee
\begin{itemize}
\item In order to determine the subgroup that is left invariant when a sextet
  develops a VEV as given in Eq.~(\ref{vev6}) we have to find the most general
  $U$ which satisfies
\be
\sum _{i=1}^3 |\,ii\,\rangle ~\rightarrow~ \sum _{i,j,k=1}^3 U_{ij}\,U_{ik}\,
|\,jk\,\rangle ~\stackrel{!}{=} ~ \sum _{i=1}^3 |\,ii\,\rangle \ .
\ee
This condition can be reformulated as
$$
\sum _{i=1}^3 U_{ij}\,U_{ik} ~=~ \sum _{i=1}^3 U^T_{ji}\,U^{}_{ik} ~=~
\delta_{jk} \ ,
$$
showing that a continuous $SO(3)$ symmetry is left unbroken by the sextet
VEV of Eq.~(\ref{vev6}). We conclude that the sextet by itself is not suitable
to break $SU(3)$ down to any of the finite groups of Table~\ref{tab-singlets}.

\item In the case of the ${\bf 10}$ we have two interesting directions. The
  VEV of Eq.~(\ref{vev10a}) is left invariant under transformations $U$ which
  satisfy 
\be
|\,123\,\rangle +\text{perm.}~\rightarrow~ \sum _{i,j,k=1}^3
U_{1i}\,U_{2j}\,U_{3k}\, |\,ijk\,\rangle + \text{perm.}
~\stackrel{!}{=} ~ |\,123\,\rangle +\text{perm.} \ .
\ee
The ten resulting conditions constrain the parameters of the $SU(3)$
transformation in Eq.~(\ref{su3trafo}). One of these conditions is obtained
from the fact that there must not be a $|\,333\,\rangle$ contribution to the
transformed state. This translates to
\be
U_{13}U_{23}U_{33} ~=~ s_{13}c_{13}^2 s_{23} c_{23} e^{-i(3\beta_1+3 \beta_2+\delta)}~=~0 \ ,
\ee
requiring $\theta_{13}=0,\frac{\pi}{2}$ or
$\theta_{23}=0,\frac{\pi}{2}$. Choosing $\theta_{13}=0$, we continue with the
condition arising from the $|\,123\,\rangle$ part of the transformed state. A
straightforward calculation yields 
\be
\cos (2\theta_{12}) \cos (2\theta_{23}) ~=~1 \ .
\ee
This can only be satisfied if both angles are either zero or
$\frac{\pi}{2}$. In that case, all remaining eight conditions are automatically
satisfied. Thus the unbroken symmetry includes a continuous phase
transformation of type $D$, see Eq.~(\ref{gen-under}), as well as $A\cdot
D$. Other elements of the unbroken group arise from setting either
$\theta_{13}=\frac{\pi}{2}$ or $\theta_{23}=0,\frac{\pi}{2}$. The resulting
unbroken group is generated by $A$ and $D$ and hence given by all elements of
the form  
\be
\left\{ D \,,\, A\cdot D \,,\, A^2 \cdot D \right\} \ ,\label{biggroup}
\ee
for all possible diagonal phase matrices $D$ with arbitrary $\vartheta_i$.
In particular the groups $\Delta(3n^2)$ and $\m Z_7 \rtimes \m Z_3$ are left
unbroken. Therefore the VEV of Eq.~(\ref{vev10a}) alone is not suitable to
break $SU(3)$ down to any of the finite groups of
Table~\ref{tab-singlets}. However, combining a ${\bf 6}$ and a ${\bf 10}$
which respectively develop VEVs in the directions of
Eqs.~(\ref{vev6},\ref{vev10a}), we end up with $\m A_4$ as the maximal
unbroken symmetry.  

The second VEV direction of interest is Eq.~(\ref{vev10b}). The corresponding
unbroken subgroup can be determined from 
\be
\sum _{i=1}^3 |\,iii\,\rangle ~\rightarrow~ \sum _{i,j,k,l=1}^3 U_{ij}\,U_{ik}\,
U_{il}\,|\,jkl\,\rangle ~\stackrel{!}{=} ~ \sum _{i=1}^3 |\,iii\,\rangle \ .
\ee
We have already seen that $\Delta(27)$ is unbroken. The question arises if
there exists a symmetry transformation $U$ which is not an element of
$\Delta(27)$. In order to find an answer we study the $|\,331\,\rangle$ and 
$|\,332\,\rangle$ contributions of the transformed state. Since both of them
must vanish, also any linear combination has to be zero. Therefore, as a
starting point, we can solve the following equation
\be
\sum _{i=1}^3 U_{i3}\,U_{i3}\,U_{i1} \,s_{12} \, e^{-i\beta_1} ~-~
\sum _{i=1}^3 U_{i3}\,U_{i3}\,U_{i2} \,c_{12} \, e^{-i\beta_2}
~ = ~ 0 \ .\label{help}
\ee
Evaluating the left-hand side leads to the condition
\be
c_{13}^2\,c_{23}\,s_{23}\,(c_{23}-e^{3i(\alpha_1+2\alpha_2)}s_{23})
~ = ~ 0 \ ,
\ee
which has solutions for $\theta_{13}=\frac{\pi}{2}$,
$\theta_{23}=0,\frac{\pi}{2}$,  as well as 
$\theta_{23}=\frac{\pi}{4}$ with 
$(\alpha_1+2\alpha_2)=\frac{2\pi }{3}\cdot\mathds{Z}$.
Each of these four cases has to be investigated using the remaining nine
conditions. Doing so it is possible to show that $\Delta(27)$ is indeed the
maximal subgroup which remains intact in this case. Hence a VEV of the form of
Eq.~(\ref{vev10b}) breaks $SU(3)$ uniquely down to $\Delta(27)$.

\item The two interesting directions of the ${\bf 15}$ are shown in
  Eqs.~(\ref{vev15a},\ref{vev15b}). They are left invariant under
  transformations which satisfy
\be
|\,12\bar 3\,\rangle +\text{perm.}~\rightarrow~ \sum _{i,j,k=1}^3
U^{}_{1i}\,U^{}_{2j}\,U^\ast_{3k}\, |\,ij\bar k\,\rangle + \text{perm.}
~\stackrel{!}{=} ~ |\,12 \bar 3\,\rangle +\text{perm.} \ ,
\ee
and 
\bea
&&\!\!\!\!\!\!\!\!\!\!
|\,11\bar 2\,\rangle \,+\,|\,22\bar 3\,\rangle \,+\,|\,33\bar 1\,\rangle 
~~\rightarrow\\
&&\!\!\!\!\!\!\!\!\!\!
 \sum _{i,j,k=1}^3
\left( U^{}_{1i}\,U^{}_{1j}\,U^\ast_{2k}+
U^{}_{2i}\,U^{}_{2j}\,U^\ast_{3k}+
U^{}_{3i}\,U^{}_{3j}\,U^\ast_{1k} 
\right)
|\,ij\bar k\,\rangle ~\stackrel{!}{=} ~ 
|\,11\bar 2\,\rangle \,+\,|\,22\bar 3\,\rangle \,+\,|\,33\bar 1\,\rangle 
 \ ,\notag
\eea
respectively. Note that the anti-triplet transforms with the complex
conjugated matrix $U^\ast$.
Following the same strategy as before, it is possible to show
that the maximal unbroken symmetries are $\m A_4$ in the case of 
Eq.~(\ref{vev15a}) as well as $\m Z_7 \rtimes \m Z_3$ for a VEV that is
aligned in the direction of  Eq.~(\ref{vev15b}).\footnote{The starting point
  in both cases is similar to Eq.~(\ref{help}). In the $\m A_4$ case one
  linearly combines the $|\,33\bar 1\,\rangle$ and $|\,33\bar 2\,\rangle$
  contributions   of the transformed state, while the $|\,13\bar 3\,\rangle$
  and $|\,23\bar 3\,\rangle$   contributions are used for $\m Z_7 \rtimes \m
  Z_3$.}   Hence depending on the VEV alignment, the ${\bf 15}$ can break
$SU(3)$ uniquely to either $\m A_4$ or $\m Z_7 \rtimes \m Z_3$. 

\end{itemize}

\section{\label{sec-pot}$\bs{SU(3)}$ invariant potentials}
\cleqn

We have seen in the previous section that certain VEV configurations of
$SU(3)$ irreps can break the continuous symmetry to a finite subgroup $\m
G$. In the following we discuss that these VEVs correspond to minima of
particular $SU(3)$ invariant scalar potentials; this exemplifies how discrete
non-Abelian symmetries can arise from the spontaneous breakdown of $SU(3)$.
As higher irreps seem to be more powerful to break $SU(3)$ uniquely to a
specific finite subgroup $\m G$, we begin our discussion with the ${\bf 15}$
which gives rise to either $\m A_4$ or $\m Z_7 \rtimes \m Z_3$. Then we consider
the irrep ${\bf 10}$ which by itself leaves the symmetry $\Delta(27)$
unbroken. Finally we also present the case of a potential that couples the
${\bf 6}$ and the ${\bf 10}$ to generate an $\m A_4$ symmetry.

\subsection{The case of a single $\bs{15}$}
Let us consider a potential with a quadratic term ${\bf 15 \otimes \ol{15}}$
as well as quartic interactions of type ${\bf 15  \otimes 15 \otimes \ol{15}
  \otimes \ol{15}}$. As the symmetric product
\be
({\bf 15  \otimes 15})_s = {\bf 6+\ol{15}+\ol{15'}+\ol{24}+\ol{60}} \
,
\ee
contains five distinct irreps, we expect five independent quartic
invariants. Therefore, the relevant potential for the ${\bf 15}$ reads
\be
V_{\bf 15}~=~ 
- m^2_{\bf 15}        \, \m I^{(0)}_{\bf 15} + 
\lambda_{\bf 15}^{} \, \m I^{(1)}_{\bf 15} +
\kappa_{\bf 15}^{}  \, \m I^{(2)}_{\bf 15} +
\rho_{\bf 15}^{}    \, \m I^{(3)}_{\bf 15} +
\tau_{\bf 15}^{}    \, \m I^{(4)}_{\bf 15} +
\eta_{\bf 15}^{}    \, \m I^{(5)}_{\bf 15} \ ,\label{pot15} 
\ee
where the invariants are obtained from different index contractions of the
tensors $T^k_{ij}$ for the ${\bf 15}$ and $\ol{T}^{ij}_k$ for the $\ol{\bf
15}$. Summing over repeated indices we define
\bea
\m I^{(0)}_{\bf 15} &=& T^k_{ij} \: \ol{T}^{ij}_k \ ,\\
\m I^{(1)}_{\bf 15} &=& T^k_{ij} \: \ol{T}^{ij}_k ~ T^l_{mn} \: \ol{T}^{mn}_l
\ ,\\
\m I^{(2)}_{\bf 15} &=& T^i_{jm} \: \ol{T}^{jn}_i ~ T^k_{ln} \: \ol{T}^{lm}_k\
,\\
\m I^{(3)}_{\bf 15} &=& T^i_{jm} \: \ol{T}^{jn}_i ~ T^m_{kl} \: \ol{T}^{kl}_n\
,\\
\m I^{(4)}_{\bf 15} &=& T^m_{ij} \: \ol{T}^{ij}_n ~ T^n_{kl} \: \ol{T}^{kl}_m\
,\\
\m I^{(5)}_{\bf 15} &=& T^i_{jm} \: T^j_{in} ~  \ol{T}^{km}_l \: \ol{T}^{ln}_k\
.
\eea
Expressing the quartic invariants in terms of the fifteen components $\Sigma_A$,
cf. Eqs.~(\ref{fifteen},\ref{fifteentensor}), we obtain polynomials of the form 
$c_{CD}^{AB} \, \Sigma_A \Sigma_B \ol{\Sigma}^C \ol{\Sigma}^D  $. It is then
straightforward to check that all five quartic invariants are linearly
independent by comparing (a subset of) the coefficients $c_{CD}^{AB}$ of these
polynomials. Eq.~(\ref{pot15}) is thus the most general potential of a
single ${\bf 15}$ with quadratic and quartic terms.

In order to see if such a potential can be minimized by the VEVs of
Eqs.~(\ref{vev15a},\ref{vev15b}), we calculate the first and second
derivatives of $V_{\bf 15}$ and insert the desired VEV alignments. In general,
setting the first derivatives to zero determines the overall scale of the
VEV in terms of the parameters of the potential, $m_{\bf 15},\lambda_{\bf
  15},\kappa_{\bf 15},\rho_{\bf 15},\tau_{\bf 15},\eta_{\bf
  15}$. Subsequently, we calculate the Hessian, i.e. the matrix of second
derivatives. A positive definite Hessian corresponds to a minimum of the
potential. Requiring positive eigenvalues then constrains the parameters of
the potential. The so obtained potential is now minimized by a VEV which
breaks $SU(3)$ down to the finite subgroup $\m G$.

Before presenting the details for the two VEV configurations of
Eqs.~(\ref{vev15a},\ref{vev15b}), a comment on the existence of zero
eigenvalues of the Hessian is in order. The potential of Eq.~(\ref{pot15}) is
symmetric under $SU(3)$ as well as a $U(1)$.\footnote{One may impose this $U(1)$
  symmetry to forbid a potential cubic term in $V_{\bf 15}$.} Both of these
symmetries are completely broken. Therefore the Hessian will automatically
have 8+1 zero eigenvalues. This means that the minimum of the potential is
assumed not only for the VEV alignments of
Eqs.~(\ref{vev15a},\ref{vev15b}) but also their $SU(3)$ 
transformed configurations. These alternative VEV alignments are still
invariant under the transformations of the finite subgroup $\m G$, however,
not in the basis of Eq.~(\ref{gen-under}) but rather 
\be
D^\prime = V D V^\dagger \ , \qquad
A^\prime = V A V^\dagger \ , \qquad
B^\prime = V B V^\dagger \ ,
\ee
where $V$ denotes the $SU(3)$ transformation to the alternative VEV
alignments.

Let us now turn to the explicit examples.
\begin{itemize}
\item Inserting the VEV alignment of Eq.~(\ref{vev15a}) into the first
  derivatives fixes the scale of the VEV to
\be
|\langle \Sigma \rangle| ~=~ \sqrt{\frac{m^2_{\bf 15}}{2\,F_{\bf 15}}} \cdot
(0,0,0,0,0,0,0,0,0,0,0,0,1,1,1)^T \ , \label{vev15mag}
\ee
with
\be
F_{\bf 15}~=~ 3\,\lambda_{\bf 15}+\kappa_{\bf 15}+\rho_{\bf 15}+\tau_{\bf 15}+\eta_{\bf 15}  \ .
\ee
As for any Higgs potential which yields a non-trivial vacuum configuration,
the coefficient $-m^2_{\bf 15}$ of the quadratic term must be negative, while
the ``effective'' coefficient $F_{\bf 15}$ of the quartic term has to be
positive. Hence we get our first conditions
\be
0\,<\,m^2_{\bf 15}  \ , \quad 0\,<\,F_{\bf 15} \ .
\ee
Additional constraints on the parameters of the potential in Eq.~(\ref{pot15})
arise from the Hessian $H$. This $30 \times 30$ matrix of second derivatives
falls into a block diagonal structure, 
\be
H~=~h_{3\times 3}^{} ~\oplus~ 3\times h_{4\times 4}^{} ~\oplus~ 3\times h_{4\times
  4}^\prime ~\oplus~ 0_{3\times 3} \ ,\label{block}
\ee
where $h_{3\times 3}^{}$ has three non-zero eigenvalues, 
\be
4 \, m^2_{\bf 15} \ , ~~~  \text{and} ~~~ 2~\times~
\frac{m^2_{\bf 15}}{F_{\bf 15}} \,(\kappa_{\bf 15} - 2 \,\eta_{\bf 15} - 2\,
\rho_{\bf 15}+ 4\,\tau_{\bf   15}) \ . \label{EV1}
\ee
The $4\times 4$ matrices $h_{4\times 4}^{}$ and $h_{4\times 4}^\prime$ both
have one zero eigenvalue as well as 
\bea
- 3\, \frac{m^2_{\bf 15}}{F_{\bf 15}} \, \eta _{\bf 15} \ ;\label{EV2}
\eea
the remaining two eigenvalues are 
\bea
\frac{m^2_{\bf 15}}{4\,F_{\bf 15}} \Big\{ 5\kappa_{\bf 15} +2\rho_{\bf 15}
  +4\tau_{\bf 15} \hspace{58mm} \label{EV3}\\
 \mp 
\sqrt{(4\tau_{\bf 15}+2\rho_{\bf 15}-3\kappa_{\bf 15})^2+16(\rho_{\bf
    15}+\kappa_{\bf 15}+2\eta_{\bf 15})^2} \Big\} , \notag 
\eea
for $h_{4\times 4}^{}$ and 
\bea
\frac{m^2_{\bf 15}}{2\,F_{\bf 15}} \Big\{ 3\kappa_{\bf 15}-5\eta_{\bf 15}
  -2\rho_{\bf 15}  +4\tau_{\bf 15}  \hspace{64.5mm} \label{EV4}\\
  \mp {\frac{1}{3}}
\sqrt{(9\eta_{\bf 15}-7\kappa_{\bf 15}+10\rho_{\bf 15}-  4\tau_{\bf
    15})^2+8(\rho_{\bf 15}+2 \kappa_{\bf 15}-4\tau_{\bf 15})^2} \Big\} , \notag 
\eea
for $h_{4\times 4}^\prime$. 

This shows that there are -- as expected -- nine zero
eigenvalues.\footnote{We have checked explicitly that the corresponding
  eigenvectors point into the directions of the $SU(3)$ and $U(1)$
  transformations.} Requiring all other eigenvalues of the Hessian to be
positive defines the set of parameters which ensures a spontaneous breaking of
$SU(3)$ to $\m A_4$. From Eq.~(\ref{EV2}) we immediately see that $\eta_{\bf
  15}<0$. The other conditions for having positive eigenvalues are less
trivial. We therefore consider the special situation in which $\lambda_{\bf
  15}=\rho_{\bf 15}=\tau_{\bf 15}=0$. In this case it is straightforward to
obtain the condition for the remaining order parameter $\kappa_{\bf 15}$; we
find 
\be
0< -\eta_{\bf 15} < \kappa_{\bf 15} \ .
\ee

\item In order to break $SU(3)$ down to $\m Z_7 \rtimes \m Z_3$ it is
  necessary to construct a potential of the type of Eq.~(\ref{pot15}) which is
  minimized by the VEV alignment of Eq.~(\ref{vev15b}). Requiring vanishing
  first derivatives sets the scale of the VEV to
\be
|\langle \Sigma^\prime \rangle| ~=~ \sqrt{\frac{m^2_{\bf 15}}{2\,F^\prime_{\bf 15}}} \cdot
(0,0,0,0,0,0,1,0,1,0,1,0,0,0,0)^T \ , \label{vev15magprime}
\ee
with
\be
F^\prime_{\bf 15}~=~ 3\,\lambda_{\bf 15}+\kappa_{\bf 15}+\rho_{\bf 15}+\tau_{\bf 15}  \ .
\ee
Both, $m^2_{\bf 15}$ and $F^\prime_{\bf 15}$ must be positive. As before, the
Hessian breaks into a block diagonal structure as given in Eq.~(\ref{block}),
with nine zero eigenvalues  corresponding to the $SU(3)$ and $U(1)$
transformations. The three eigenvalues of $h_{3\times 3}$ read
\be
4 \, m^2_{\bf 15} \ , ~~~  \text{and} ~~~ 2~\times~
\frac{m^2_{\bf 15}}{F^\prime_{\bf 15}} \,(4\kappa_{\bf 15}  - 2\,
\rho_{\bf 15}+ 4\,\tau_{\bf   15}) \ . \label{EV1prime}
\ee
The submatrices $h_{4\times 4}$ and $h^\prime_{4\times 4}$ turn out to be
identical up to a trivial sign change,
\be
h_{4\times 4} ~=~ \mathrm{Diag}(1,1,-1,-1) \cdot  h^\prime_{4\times 4}
\cdot \mathrm{Diag}(1,1,-1,-1) \ ,
\ee
so that their eigenvalues are identical. One of the four eigenvalues is always
zero while, in general, the other three eigenvalues $x_i$ are
non-vanishing. They can be determined as the solutions to the following cubic
polynomial 
\bea
&&4  \xi_i^3
- \xi_i^2 \left(  18 \eta_{\bf 15}^{}  
+ 7 \kappa_{\bf 15}^{}  
- 2 \rho_{\bf 15}^{} 
+ 12 \tau_{\bf 15}^{} \right)  \notag \\
&&+ \xi_i \left( 18 \eta_{\bf 15}^2
+ 27 \eta_{\bf 15}^{}  \kappa_{\bf 15}^{}  
- 18 \eta_{\bf 15}^{}  \rho_{\bf 15}^{}  
- 5 \rho_{\bf 15}^2   
+ 36 \eta_{\bf 15}^{}  \tau_{\bf 15}^{}    
+ 20 \kappa_{\bf 15}^{}  \tau_{\bf 15}^{} \right)  \notag \\
&&- 3 \left( 8 \eta_{\bf 15}^2 \kappa_{\bf 15}^{} 
- 8 \eta_{\bf 15}^2 \rho_{\bf 15}^{}  
- 3 \eta_{\bf 15}^{}  \rho_{\bf 15}^2 
+ 8 \eta_{\bf 15}^2 \tau_{\bf 15}^{}  
+ 12 \eta_{\bf 15}^{}  \kappa_{\bf 15}^{}  \tau_{\bf 15}^{} \right) ~=~0 \ ,~~~~~~
\eea
where $\xi_i = \frac{F^\prime_{\bf 15}}{m^2_{\bf 15}} x_i$. Note that $\xi_i$
and $x_i$ have identical signs. To present a scenario in which all
non-vanishing eigenvalues of the Hessian are positive let us again consider
the special case with $\lambda_{\bf 15}=\rho_{\bf   15}=\tau_{\bf 15}=0$. The
condition $0<F^\prime_{\bf 15}$ as well as Eq.~(\ref{EV1prime}) demand
positive $\kappa_{\bf 15}$ in that case. With these assumptions the cubic
polynomial simplifies and we can calculate the three roots. However, as the
analytic expressions are rather lengthy, we show the results graphically in
Figure~\ref{T7-plot}. In order to have a minimum all three eigenvalues must
be positive. This immediately implies positive $\eta_{\bf 15}$. 
So in the case where $\lambda_{\bf 15}=\rho_{\bf   15}=\tau_{\bf 15}=0$, the
conditions to get a VEV that breaks $SU(3)$ down to $\m Z_7\rtimes \m Z_3$ are 
\be
0\,<\,\kappa_{\bf 15} \ , \quad0\,<\,\eta_{\bf 15} \ .
\ee

\begin{figure}
\begin{center}
\includegraphics[width=8.5cm]{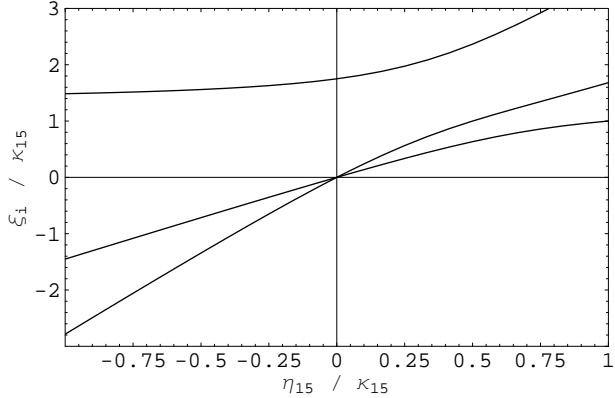} \qquad
\end{center}
\caption{\label{T7-plot}$\m Z_7\rtimes \m Z_3$ from the ${\bf 15}$ of $SU(3)$:
  the three non-vanishing scaled eigenvalues $\frac{\xi_i}{\kappa_{\bf 15}}$ of
    the sub-Hessian $h_{4\times 4}$ are shown as functions of $\frac{\eta_{\bf
        15}}{\kappa_{\bf 15}}$ in the case where $\lambda_{\bf 15}=\rho_{\bf
      15}=\tau_{\bf 15}=0$.}   
\end{figure}

\end{itemize}

\subsection{The case of a single $\bs{10}$}

Similar to the previous case, we consider a potential of a single ${\bf 10}$
which has a mass term ${\bf 10 \times \ol{10}}$ as well as quartic
interactions of type  ${\bf 10 \times 10 \times \ol{10}\times \ol{10}}$. The
symmetric product 
\be
({\bf 10 \times 10})_s \,=\,  
{\bf 27 + 28} \ ,
\ee
shows that we can only write down two independent quartic $SU(3)$
invariants. Hence, the potential for the ${\bf 10}$ takes the form
\be
V_{\bf 10}~=~ 
- m^2_{\bf 10}        \, \m I^{(0)}_{\bf 10} + 
\lambda_{\bf 10}^{} \, \m I^{(1)}_{\bf 10} +
\kappa_{\bf 10}^{}  \, \m I^{(2)}_{\bf 10}  \ ,\label{pot10} 
\ee
with
\bea
\m I^{(0)}_{\bf 10} &=& T^{}_{ijk} \: \ol{T}^{ijk}_{} \ ,\label{inv10=0}\\
\m I^{(1)}_{\bf 10} &=& T^{}_{ijk} \: \ol{T}^{ijk}_{} ~ T^{}_{lmn} \: \ol{T}^{lmn}_{}
\ ,\label{inv10=1}\\
\m I^{(2)}_{\bf 10} &=& T^{}_{ijm} \: \ol{T}^{ijn}_{} ~ T^{}_{kln} \: \ol{T}^{klm}_{}\
.\label{inv10=2}
\eea
Using the VEV configuration of Eq.~(\ref{vev10b}) which breaks $SU(3)$
uniquely down to $\Delta(27)$, we can determine the scale of the VEV alignment
by setting the first derivatives to zero. We obtain
\be
|\langle \psi \rangle| ~=~ \sqrt{\frac{m^2_{\bf 10}}{2\,F_{\bf 10}}} \cdot
(1,1,1,0,0,0,0,0,0,0)^T \ , \label{vev10mag}
\ee
with
\be
F_{\bf 10}~=~ 3\,\lambda_{\bf 10}+\kappa_{\bf 10}  \ .
\ee
Having a minimum requires positive values for $m^2_{\bf 10}$ and $F_{\bf
  10}$. The other constraints on the parameters of the potential arise from
the Hessian. The $20 \times 20$ matrix can be calculated analytically,
yielding eleven zero eigenvalues as well as 
\be
4m_{\bf 10}^2 \ ,~~~ ~  
6 ~\times ~  \frac{4 m_{\bf 10}^2 \kappa^{}_{\bf 10}}{3 F^{}_{\bf 10}}\
,~~~ \mathrm{and} ~~~   
2 ~\times ~  \frac{4 m_{\bf 10}^2 \kappa^{}_{\bf 10}}{F^{}_{\bf 10}}\ .
\ee
Consequently, we need positive $\kappa_{\bf 10}$ in order to have a potential
which is minimized by the VEV alignment of Eq.~(\ref{vev10b}). The number of
zero eigenvalues of the Hessian can be understood as follows.
Eight zeros are due to the eight broken generators of $SU(3)$; another zero
eigenvalues arises because the VEV also breaks a global $U(1)$. The remaining
two vanishing eigenvalues are related to the existence of the second
$\Delta(27)$ singlet within the ${\bf 10}$. Any linear combination of the VEV
alignments in Eq.~(\ref{vev10a}) and Eq.~(\ref{vev10b}) leaves the group
$\Delta(27)$ intact. Hence, the additional two zero eigenvalues of the Hessian
correspond to the directions of the real and the imaginary part of $\psi_{10}$. 
Sliding along this direction, the residual symmetry will remain $\Delta(27)$ as
long as $\langle \psi_{1,2,3} \rangle \neq 0$. Only in the special vacuum
where the first three components of the ${\bf 10}$ vanish identically, we end
up with the bigger group given in Eq.~(\ref{biggroup}). This can be avoided
by small deformations of the potential. A simple scenario could consist in
adding a second ${\bf 10}$ which is aligned as in Eq.~(\ref{vev10a}),
cf. section~\ref{sec10and6}. We can then introduce a quartic term which
couples the two different ${\bf 10}$s as follows,
\be
 \sum_{a,b=1}^{10} (\psi^{}_a \, \ol{\psi_{a}^{\prime}}) \, (\ol{\psi_{b}^{}} \, \psi^{\prime}_b
) \ .
\ee
Note that such a term is always positive or zero. Assuming this term to enter
the potential with a positive coupling constant, the minimum arises if
$\sum_{a=1}^{10} \langle \psi^{}_a \rangle \, \langle \ol{\psi_{a}^{\prime}}
\rangle = 0$. With $\langle \psi^\prime_a\rangle=~\!\!0$ for $a=1,2,...,9$, this
entails vanishing $\langle \psi_{10} \rangle$. Therefore, the VEV of
$\psi$ is driven to the alignment of Eq.~(\ref{vev10b}) which breaks $SU(3)$
uniquely down to $\Delta(27)$.

\subsection{\label{sec10and6}The case of a $\bs{10}$ and a $\bs{6}$}

We have seen in section~\ref{sec-subgroup} that the combination of a ${\bf 6}$
and a ${\bf 10}$ with alignments along the directions of
Eqs.~(\ref{vev6},\ref{vev10a}) gives rise to a residual $\m A_4$
symmetry. In the following we show that there exists a potential which assumes
its minimum for exactly these VEV alignments. The most general renormalizable
potential of one ${\bf 6}$ and one ${\bf 10}$ consists of thirteen
invariants. It reads
\bea
V_{\bf 6+10}&=&
- m^2_{\bf 6}        \, \m I^{(0)}_{\bf 6} 
+ \lambda_{\bf 6}^{} \, \m I^{(1)}_{\bf 6} 
+ \kappa_{\bf 6}^{}  \, \m I^{(2)}_{\bf 6}  
+ \rho_{\bf 6}^{}  \, \m I^{(3)}_{\bf 6}  \notag \\
&& 
- m^2_{\bf 10}        \, \m I^{(0)}_{\bf 10} 
+ \lambda_{\bf 10}^{} \, \m I^{(1)}_{\bf 10} 
+ \kappa_{\bf 10}^{}  \, \m I^{(2)}_{\bf 10}  
+ \rho_{\bf 10}^{}  \, \m I^{(3)}_{\bf 10} 
+ \tau_{\bf 10}^{}  \, \m I^{(4)}_{\bf 10}  \notag \\
&&
+ \, \eta_{1}^{} \, \m I^{(1)}_{\bf 6+10} 
+ \eta_{2}^{}  \, \m I^{(2)}_{\bf 6+10}  
+ \eta_{3}^{}  \, \m I^{(3)}_{\bf 6+10} 
+ \eta_{4}^{}  \, \m I^{(4)}_{\bf 6+10} \ ,\label{pot6and10} 
\eea
with 
\bea
\m I^{(0)}_{\bf 6} &=& T^{}_{ij} \: \ol{T}^{ij}_{} \ ,\\
\m I^{(1)}_{\bf 6} &=& T^{}_{ij} \: \ol{T}^{ij}_{} ~ T^{}_{kl} \: \ol{T}^{kl}_{}
\ ,\\
\m I^{(2)}_{\bf 6} &=& T^{}_{ik} \: \ol{T}^{il}_{} ~ T^{}_{jl} \:
\ol{T}^{jk}_{}\ ,\\
\m I^{(3)}_{\bf 6} &=& \epsilon^{ijk} \, T^{}_{1i} \: {T}_{2j}^{} \:
T^{}_{3k} ~+~ \mathrm{h.c.} \ ,
\eea
\bea
\m I^{(3)}_{\bf 10} &=& \epsilon^{xx'k} \epsilon^{yy'l} \, T^{}_{ixy} \:
T^{}_{jx'y'} ~ T^{}_{mkl} \: \ol{T}^{ijm}_{}~+~ \mathrm{h.c.}\ ,\\
\m I^{(4)}_{\bf 10} &=& \epsilon^{xx'k} \epsilon^{yy'l} \, T^{}_{ixy} \:
T^{}_{jx'y'} ~ \epsilon^{vv'i} \epsilon^{ww'j} \, T^{}_{kvw} \: T^{}_{lv'w'}~+~ \mathrm{h.c.}\
,
\eea
\bea
\m I^{(1)}_{\bf 6+10} &=& T^{}_{ij} \: \ol{T}^{ij}_{} ~ T^{}_{klm} \:
\ol{T}^{klm}_{} \ ,\\
\m I^{(2)}_{\bf 6+10} &=& T^{}_{ijm} \: \ol{T}^{ij}_{} ~ T^{}_{kl} \:
\ol{T}^{klm}_{} \ ,\\
\m I^{(3)}_{\bf 6+10} &=& T^{}_{ijm} \: \ol{T}^{ijn}_{} ~ T^{}_{kn} \:
\ol{T}^{km}_{} \ ,\\
\m I^{(4)}_{\bf 6+10} &=& \epsilon^{xx'k} \epsilon^{yy'l} \, T^{}_{ixy} \:
T^{}_{jx'y'} ~ T^{}_{kl} \: \ol{T}^{ij}_{}~+~ \mathrm{h.c.} \ ,
\eea
and $\m I^{(0)}_{\bf 10}$, $\m I^{(1)}_{\bf 10}$, $\m I^{(2)}_{\bf 10}$ as
given in Eqs.~(\ref{inv10=0}-\ref{inv10=2}). The tensors $T_{...}$ with three
indices correspond to the ${\bf 10}$ while those with two indices stand for
the ${\bf 6}$; a bar indicates complex conjugate
representations. $\epsilon^{ijk}$ denotes the totally antisymmetric tensor
with $\epsilon^{123}=1$. Note that all invariants which contain this $\epsilon$
tensor are {\it not} symmetric under a general $U(1)$ while all other invariants
feature such a $U(1)$ symmetry.

\begin{figure}
\begin{center}
\includegraphics[width=8.5cm]{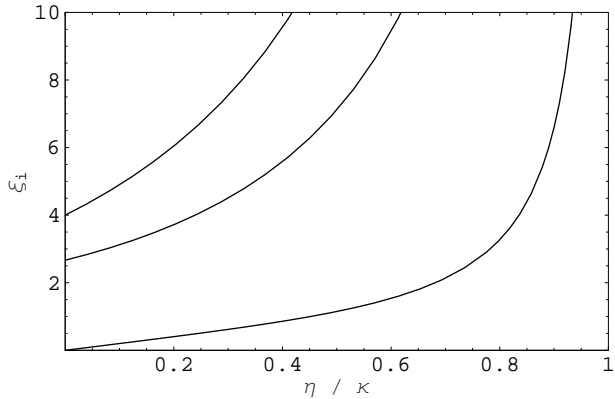} \qquad
\end{center}
\caption{\label{a4special-plot}$\m A_4$ from a ${\bf 6}$ and a ${\bf 10}$ of $SU(3)$:
  the three non-vanishing scaled eigenvalues $\xi_i$ of
    the sub-Hessian $h_{4\times 4}^{\prime}$ are shown as functions of 
 $\frac{\eta}{\kappa}$ in the special case of 
Eqs.~(\ref{a4-special1},\ref{a4-special2}).}   
\end{figure}

Evaluation of the first derivatives using the alignment directions 
of Eqs.~(\ref{vev6},\ref{vev10a}) fixes the scale of the VEVs,
\bea
\langle \chi \rangle \,=\, R_{\bf 6} \,
(1,1,1,0,0,0)^T \ , \qquad
\langle \psi \rangle \,=\, R_{\bf 10} \,
(0,0,0,0,0,0,0,0,0,1)^T \ .
\label{vev6and10mag}
\eea
Despite the lack of a general $U(1)$ symmetry we can assume real VEVs $R_{\bf
6}$ and $R_{\bf 10}$ for our purposes, because any potential $V'$ which is
minimized by complex VEVs corresponds to a modified potential $V$ in which the
coupling constants absorb the phases of the complex VEVs, thus rendering the
latter real. With this assumption we obtain the following two conditions on
$R_{\bf 6}$ and $R_{\bf 10}$,
\bea
0&=& -3 m_{\bf 6}^2 
+  R_{\bf 10}^{2} (3 \eta_1^{} + \eta_3^{} - 2 \eta_4^{})
+ 3 R_{\bf 6}^{} ( 6 R_{\bf 6}^{} \lambda_{\bf 6}^{} 
                 + 2 R_{\bf 6}^{}  \kappa_{\bf 6}^{}  + \rho_{\bf 6}^{}) 
\ ,\notag\\
0&=& -3 m_{\bf 10}^2 
+ 3 R_{\bf 6}^2 ( 3 \eta_1^{} + \eta_3^{} - 2 \eta_4^{})  
+ 2  R_{\bf 10}^2 (3\lambda_{\bf 10}^{}+\kappa_{\bf 10}^{}
                     +2\rho_{\bf 10}^{}+4\tau_{\bf 10}^{}) 
\ .\notag
\eea
For the sake of simplicity we assume $\rho_{\bf 6}^{} =0$.\footnote{This could
be enforced by a $U(1)$ symmetry under which the ${\bf 6}$ carries
non-vanishing charge while the ${\bf 10}$ is neutral.} Then the above
conditions are satisfied for
\bea
R_{\bf 6}^{2} &=&\frac{2m_{\bf 6}^2 (3\lambda_{\bf 10}^{}+\kappa_{\bf 10}^{}
 +2\rho_{\bf 10}^{}+4\tau_{\bf 10}^{})-m_{\bf 10}^2 (3 \eta_1^{} + \eta_3^{} - 2 \eta_4^{}) }{4(3\lambda_{\bf 6}^{}+\kappa_{\bf
   6}^{} ) (3\lambda_{\bf 10}^{}+\kappa_{\bf 10}^{}
                     +2\rho_{\bf 10}^{}+4\tau_{\bf 10}^{})
-(3 \eta_1^{} + \eta_3^{} - 2 \eta_4^{})^2 }
\ , \\[1mm]
R_{\bf 10}^{2}&=&\frac{6m_{\bf 10}^2 (3\lambda_{\bf 6}^{}+\kappa_{\bf 6}^{}
)-3 m_{\bf 6}^2 (3 \eta_1^{} + \eta_3^{} - 2 \eta_4^{}) }{4(3\lambda_{\bf 6}^{}+\kappa_{\bf
   6}^{} ) (3\lambda_{\bf 10}^{}+\kappa_{\bf 10}^{}
                     +2\rho_{\bf 10}^{}+4\tau_{\bf 10}^{})
-(3 \eta_1^{} + \eta_3^{} - 2 \eta_4^{})^2 }
\ .
\eea
Evaluating the second derivatives for these VEVs yields a block diagonal
structure for the $32\times 32$ Hessian
\be
H~=~h_{1\times 1}^{} 
~\oplus~h_{4\times 4}^{} 
~\oplus~ 3\times h_{4\times 4}^{\prime} 
~\oplus~ 3\times h_{4\times 4}^{\prime\prime} 
~\oplus~ 0_{3\times 3} \ .\label{block32}
\ee
In general, $h_{1\times 1}^{}$ and $h_{4\times 4}^{}$ have no vanishing
eigenvalue, while $h_{4\times 4}^{\prime}$ and $h_{4\times 4}^{\prime\prime}$
each have one zero eigenvalue. Therefore the full Hessian exhibits nine zero
eigenvalues corresponding to the directions of the eight $SU(3)$
transformations plus an extra $U(1)$ transformation. Notice that there exists
only one $U(1)$ symmetry and not two because the charge of the ${\bf 10}$ is
fixed to be neutral. In order to have a minimum we need the remaining 23
eigenvalues to be positive. This constrains the set of parameters of the
potential $V_{\bf 6+10}$ in Eq.~(\ref{pot6and10}). As an example we discuss
the special case where 
\be
m_{\bf 6} = m_{\bf 10} = m\ , \qquad 
\kappa_{\bf 6} = \kappa_{\bf 10} = \kappa\ , \qquad
\eta_4 = \eta \ ,
\label{a4-special1}
\ee
\be
\lambda_{\bf 6}=\lambda_{\bf 10} = \rho_{\bf 6}=\rho_{\bf 10}=\tau_{\bf 10}=
\eta_1=\eta_2=\eta_3= 0 \ .\label{a4-special2}
\ee
Then the VEVs simplify to
\be
R_{\bf 6}^2 = \frac{m^2}{2(\kappa-\eta)} \ , \qquad
R_{\bf 10}^2 = \frac{3m^2}{2(\kappa-\eta)} \ ,
\ee
requiring positive $m^2$ as well as $\eta<\kappa$. The eigenvalues of the
sub-Hessians are calculated to be
\bea
h_{1\times 1}^{}: && \frac{4m^2 \eta}{\kappa-\eta} \ ,\\
h_{4\times 4}^{}: && 4m^2 \ ,
\qquad \frac{4m^2 (\kappa+\eta)}{\kappa-\eta}
\qquad 2~\times~ \frac{4m^2 \kappa}{\kappa-\eta} \ ,\\[2mm]
h_{4\times 4}^{\prime}: &&  x_1 \ , \quad x_2 \ , \quad x_3 \ , \quad 0
\ , \\[2mm]
h_{4\times 4}^{\prime\prime}: && \frac{4m^2 (2\kappa+3\eta)}{3(\kappa-\eta)}\ ,
\qquad 
\frac{m^2 \eta(13\pm\sqrt{109})}{3(\kappa-\eta)} \ , \qquad 0 \ ,
\eea
where $x_i$ are the solutions to the cubic polynomial
\bea
&& 3 \xi_i^3 (\eta-\kappa)^3 +2 \xi^2_i (\eta-\kappa)^2 (11\eta+10\kappa)
\notag\\ 
&&+4 \xi_i^{} (\eta-\kappa) (7\eta^2+22\eta\kappa+8\kappa^2)
-16\eta (\eta^2-2\eta\kappa-4\kappa^2) ~=~ 0 \ ,~~~~~~~
\eea
with $\xi_i=\frac{x_i}{m^2}$. Figure~\ref{a4special-plot} presents the results
graphically for the relevant region 
\be
0\,<\,\eta\,<\,\kappa \ ,
\ee
which is obtained from requiring positive values for the other eigenvalues of
the Hessian. From this example it is clear that parameter ranges exist in which
the potential $V_{\bf 6+10}$ of Eq.~(\ref{pot6and10}) is minimized by the
alignments of Eqs.~(\ref{vev6},\ref{vev10a}). Hence $\m A_4$ can result as the
discrete remnant of a spontaneously broken $SU(3)$ symmetry.


\section{\label{sec-conclusion}Conclusion}
\cleqn

In this paper we have investigated the possibility of obtaining a non-Abelian
discrete family symmetry $\m G$ from an underlying $SU(3)$ gauge symmetry. 
Such a scenario is appealing in the sense that the residual discrete symmetry
is protected against violations by quantum gravity effects. We have first
identified the higher $SU(3)$ representations which contain singlets under
various discrete subgroups. These are potential candidates of fields
that are capable of breaking $SU(3)$ down to $\m G$. Fixing the basis of the
subgroup, we have determined the $\m G$ singlet directions and checked whether
these vacuum alignments leave invariant the desired subgroups or something
bigger. Scrutinizing various $SU(3)$ invariant potentials which involve higher
representations comprises the central part of the paper. Constraining
ourselves to the irreps ${\bf 6}$, ${\bf 10}$ and ${\bf 15}$ we found that $\m
A_4$, undoubtedly the most popular family symmetry, can be generated from
either a single ${\bf 15}$ or alternatively a combination of a ${\bf 6}$ and a
${\bf 10}$. Similarly, the group $\m Z_7 \rtimes \m Z_3$ is obtained from a
single ${\bf 15}$, however using different numerical values for the coupling
constants of the potential. Finally, a single ${\bf 10}$ allows to break
$SU(3)$ down to the group $\Delta(27)$. These results show that an $SU(3)$
gauge symmetry can give rise to non-Abelian discrete family symmetries,
sometimes adopting only one $SU(3)$ breaking multiplet. 
Having discussed the above examples in great detail, it should be clear how to
proceed in the case of other discrete symmetries $\m G$.
For instance, the family symmetry $\m{PSL}_2(7)$ is expected to arise from
an appropriate vacuum alignment of the ${\bf 15'}$ of $SU(3)$. This case will
be treated elsewhere. In the context of a concrete model \cite{King:2009tj}
we hope to find a solution to an unexplained tuning which is required to
generate the correct vacuum structure of the flavon sextets.

We conclude by pointing out that our work does not address the question of
how the breaking of the continuous symmetry is communicated to the Yukawa
sector. In general this is a very model dependent problem as there are
different choices for assigning the Standard Model fermions as well as the
$\m G$ breaking flavons to irreps of the underlying $SU(3)$ symmetry. 
Depending on this choice the product rules constrain the allowed
interactions of the $SU(3)$ breaking field(s) to the chiral fermions and
flavons. Such an investigation should be carried out within the context of a
specific flavor model and is therefore beyond the scope of our paper.


\section*{Acknowledgments}

I am indebted to Pierre Ramond and Steve King for encouragement
and stimulating discussions. I wish to specially thank Pierre Ramond 
for reading the manuscript and his helpful comments. This work is supported by
the STFC Rolling Grant ST/G000557/1.




\end{document}